\documentclass[12pt,a4paper]{article}
\usepackage[T2A]{fontenc}
\usepackage[cp1251]{inputenc}
\usepackage{cite}
\usepackage{mathtext}
\usepackage{amssymb,amsthm,amsmath}
\usepackage{array}
\usepackage[dvips]{epsfig}

\begin{document}
\author{\bf Yu.A. Markov$\!\,$\thanks{e-mail:markov@icc.ru}
$\,$, M.A. Markova$^*$}
\title{On the fluctuation-dissipation theorem\\
for soft fermionic excitations in a hot QCD plasma}
\date{\it Institute for System Dynamics\\
and Control Theory, Siberian Branch\\
of Academy of Sciences of Russia,\\
P.O. Box 1233, 664033 Irkutsk, Russia}

\thispagestyle{empty}
\maketitle{}
\def\theequation{\arabic{section}.\arabic{equation}}

\[
{\bf Abstract}
\]
We discuss two ways of deriving the fluctuation-dissipation theorem (FDT) for soft fermion excitations in a hot non-Abelian plasma
being in a thermal equilibrium. The first of them is based on the extended (pseudo)classical model in describing a quark-gluon plasma
suggested in [Yu.A. Markov, M.A. Markova, Nucl. Phys. A {\bf 784} (2007) 443], while the second one rests on the standard technique of 
calculation of the FDT for thermodynamically equilibrium systems. We show that full accounting all subtleties that are common to the 
fermion system under consideration, results in perfect coincidence of thus obtained FDTs. This provides a rather strong argument for 
the validity of the pseudoclassical model suggested.
{}


\newpage

\section{Introduction}
\setcounter{equation}{0}

In our two previous papers \cite{markov_NPA_2007, markov_IJMPA_2010} we have proposed an approach to construction of the
generalized (pseudo)classical theory for a unified description of a broad spectrum of interaction processes of soft and hard quark-gluon
plasma excitations obeying both Fermi and Bose statistics. Setting into consideration Grass\-mann (anticommuting) color charges
$\theta = (\theta^i)$ and $\theta^{\dagger}=(\theta^{\dagger i}),\,i=1,\ldots,N_c$, along with usual (commuting) classical
color charge $Q=(Q^a),\, a=1,\ldots,N_c^2-1$, of hard particle, is a principle feature of this extended theory. A number of examples
of an explicit constructing the probabilities for various scattering processes involving the bosonic and fermionic hard and
soft modes, has been given. Also possible applications of the theory we have developed to an important problem of the calculation
of energy losses of fast Yang-Mills particles passing through a hot QCD matter, were considered.\\
\indent
However, in spite of the reasonableness of presuppositions in the construction of the generalized theory mentioned above and
also obtaining rather nontrivial results within the limits of this theory, we would like to have independent, physically transparent
and simultaneously a simple way for critical examination of the approach suggested in \cite{markov_NPA_2007, markov_IJMPA_2010}.
This independent testing must be founded on some fundamental, i.e. model-independent, physical principle.
As such a principle in the present work we have chosen the {\it fluctuation-dissipation theorem} which as is
generally known \cite{kubo_1966}, admits an abstract formulation convenient for applications of this theorem to fields
of equilibrium thermal fluctuations of any physical nature: electromagne\-tic, mechanical, temperature, entropy and so on.
The FDT already for a long time, is an integral part in courses on the statistical mechanics
\cite{zubarev_book_1974, kubo_book_1985, akhiezer_book_1986, klimontovich_book_1986, sethna_book_2006, schwabl_book_2008}
(see, also recent review \cite{marconi_2008}).\\
\indent
A distinctive characteristic of the problem under consideration, is the fact that an external perturbation described by a Hamiltonian
$\hat{H}_t^1$, changes the number of particles and antiparticles in the system, i.e., in other words,
an operator of the total number of particles $\hat{N}$ (more precisely, the difference between the number of particles and antiparticles)
does not commute with the $\hat{H}_t^1$ (but it commutes with a Hamiltonian of the many-particle system $\hat{H}_0$). And though
such a case has already been discussed in the general statement in textbook materials (see, e.g., \cite{akhiezer_book_1986, schwabl_book_2008}),
nevertheless as far as we know, any specific physical situation, where this circumstance would play crucial role, has not been considered.
As a consequence of this fact, a concrete expression for the fluctuation-dissipation theorem in which this noncommutativity could be manifested
by obvious fashion, has not been given anywhere. For this reason, in the present paper we have found pertinent to give
the comprehensive derivation of the FDT relations for the physical problem at hand.\\
\indent
The paper is organized as follows. In Section 2, by using a simple classical model
of the quark-gluon plasma, we derive the spectral density for thermal bosonic fluctuations.
We show that the expression obtained exactly coincides with that following from the corresponding
fluctuation-dissipation theorem. Section 3 is concerned with deriving relevant spectral density for
thermal fluctuations of the quark-antiquark field in the medium within the extended (pseudo)classical model of
the QGP. Sections 4 and 5 are devoted to determining the FDT relations for the volume field
density of the generalized forces and the stationary and homogeneous soft fields of the system in
question, obeying Fermi-Dirac statistics. In the former section we obtain the FDT for fluctuations
of the hard sources $\eta_{\alpha}^i$ and $\bar{\eta}_{\alpha}^i$, while in the latter a similar theorem is derived
for fluctuations of the soft spinor fields $\psi_{\alpha}^i$ and  $\bar{\psi}_{\alpha}^i$. At the end
of Section 5 we discuss the condition of matching two expressions for the soft-quark spectral density obtained
within the pseudoclassical model of QGP and the fluctuation-dissipation theorem.

\section{Correlation function for soft Bose-excitations}
\setcounter{equation}{0}

Before we proceed to calculation of the spectral density for fermionic fluctuations and to derivation of the relevant fluctuation-dissipation theorem, we would like to
consider briefly similar questions for much-studied case of fluctuations of a gauge field in the medium. There are no new results to be reported here. The main goal
of this section is to outline the principle points of obtaining the FDT for soft bosonic fluctuations. Further, in subsequent sections we shall follow these points in deriving
a similar FDT for the fermion degree of freedom of a non-Abelian plasma. A somewhat different approach to obtaining the fluctuation-dissipation theorem for bosonic sector of plasma
fluctuations can be found in the papers by D.F. Litim and C. Manuel \cite{litim_1999}.\\
\indent
As a first step we shall calculate the spectral density for bosonic fluctuations in the hot non-Abelian plasma within the classical model
suggested by U. Heinz more than 20 years ago \cite{heinz_1983, heinz_1985, heinz_1986}. The second step is a comparison of the obtained expression
for the spectral density with a similar expression resulting from the FDT for fluctuations of a gauge field.\\
\indent
According to \cite{heinz_1985}, the soft gluon field $A_{\mu}^a$ induced by a hard test particle (which is centered at the position ${\bf x}_0$) in the momentum representation is
\begin{equation}
A^{a}_{\mu}(k) = -\,^{\ast}{\cal D}_{\mu\nu}(k)
j^{a\nu}(k;{\bf x}_{0}),
\quad k\equiv(\omega,{\bf k})
\label{eq:2q}
\end{equation}
where in turn the current of the hard color-charged particle reads
\begin{equation}
j^{a\nu}(k;{\bf x})=
\frac{\,g}{(2\pi)^3}\;v^{\nu}
Q^a\delta(v\cdot k)\,
{\rm e}^{-i{\bf k}\cdot\,{\bf x}_{0}},\quad
v=(1,{\bf v}).
\label{eq:2w}
\end{equation}
Here, $\!\,^{\ast}{\cal D}_{\mu\nu}(k)$ is the gluon propagator in the hard thermal loop (HTL) approximation\footnote{\,In the high-temperature QCD within the hard thermal loop approximation
it is customary to denote all the effective (i.e. HTL-resummed) propagators and vertices by $^{\ast}{\cal D}$, $^{\ast}\!S$, $^{\ast}{\Gamma}$ and so on, with the aim to distinguish them from the
exact quantities ${\cal D}$, $S$, $\Gamma, \ldots$ (see, for example, the review \cite{blaizot_2002}). We shall maintain these notions throughout our paper. As usually we put the sign of complex conjugation ``$\,{\ast}\,$'' to the right of a quantity.}; ${\bf v}$ and $Q^a$ are the velocity and classical color charge of the hard particle, respectively. In the simplest approximation
the last two quantities are taken time-independent ones.\\
\indent
Further, we consider the correlation function $\left\langle A_{\mu}^{\ast a}(k)A_{\nu}^b(k^{\prime})\right\rangle$ with gauge
field (\ref{eq:2q}). We need first to define what is meant by the angular brackets of averaging. The average procedure involves
the integration over all possible initial positions ${\bf x}_0$ of the hard particle. Then, the average involves the integration
with respect to color charge with the measure $dQ$. An explicit form of the latter is written out in Appendix A. Besides,
the average over the distributions of hard particles: quarks $(\textmd{Q})$, antiquarks $(\bar{\textmd{Q}})$ and gluons $(\textmd{G})$
being in thermal equilibrium, should be added. Thus we can write
\begin{equation}
\begin{split}
\left\langle A_{\mu}^{\ast a}(k)A_{\nu}^b(k^{\prime})\right\rangle&=
2\,\frac{g^2}{(2\pi)^6}\!\sum\limits_{\;\zeta=\,\textmd{Q},\,\bar{\textmd{Q}},\,\textmd{G}\,}\int\!dQ\,Q^aQ^{b\!}\!
\int\!d{\bf x}_{0}\,{\rm e}^{i({\bf k}-{\bf k}^{\prime})\cdot\,{\bf x}_{0}\!}
\!\int\!{\bf p}^2 f_{|{\bf p}|}^{(\zeta)}\,\frac{d|\,{\bf p}|}{2\pi^2}\\
&\times\left(\!\,^{\ast}{\cal D}_{\mu\mu^{\prime}}(k)\right)^{\ast}
\left(\!\,^{\ast}{\cal D}_{\nu\nu^{\prime}}(k^{\prime})\right)\!
\int\!\frac{d\Omega_{{\bf v}}}{4\pi}\,v^{\mu^{\prime}}v^{\nu^{\prime}}
\delta(v\cdot k)\delta(v\cdot k^{\prime}).
\end{split}
\label{eq:2e}
\end{equation}
The overall factor 2 takes into account that hard gluons and massless hard (anti)quarks have two helicity states.
The integration over ${\bf x}_0$ is trivially performed
\begin{equation}
\int\!d{\bf x}_{0}\,{\rm e}^{i({\bf k}-{\bf k}^{\prime})\cdot\,{\bf x}_{0}}=
(2\pi)^3\delta({\bf k}-{\bf k}^{\prime}).
\label{eq:2r}
\end{equation}
Furthermore, considering the last expression, in the second line of (\ref{eq:2e}), we have
\begin{equation}
\delta(v\cdot k)\delta(v\cdot k^{\prime})=\delta(\omega-\omega^{\prime})
\delta(\omega-{\bf v}\cdot{\bf k}).
\label{eq:2t}
\end{equation}
The integral over the color charge is defined by a formula
\begin{equation}
\int\!dQ\,Q^aQ^b\!=\left(\frac{C_2^{(\zeta)}}{d_A}\right)\delta^{ab},\quad d_A=N_c^2-1.
\label{eq:2y}
\end{equation}
The exact definitions of the constants $C_2^{(\zeta)}$ in terms of the group invariants, are given in Appendix A.\
\indent
It takes a little more efforts to analyze the integral over a solid angle. Let us present it
in the form of an expansion in terms of the basis of Lorentz-covariant matrices fixed in the rest frame of the medium
\[
\int\!\frac{d\Omega_{{\bf v}}}{4\pi}\,v^{\mu}v^{\nu}\delta(v\cdot k)=
a_tP^{\mu\nu}(k)+a_l\,Q^{\mu\nu}(k)+\ldots,
\]
where $a_{t,l}$ are unknown for a while constants; $P^{\mu\nu}$ and $Q^{\mu\nu}$ are transverse and longitudinal projectors
whose explicit forms are given in Appendix B. The dots designates contributions of the other possible tensor structures.\\
\indent
Let us introduce the spectral density for soft bosonic fluctuations through a relation\footnote{\label{foot_1}\,On the right-hand side of
the definition of the spectral density (\ref{eq:2u}), the factor $(2\pi)^4$ is absent since we use the following formulae
for the direct and inverse Fourier transforms:
\[
F(x)=\!\int\!\tilde{F}(k)\,{\rm e}^{-ik\cdot x}dk,\quad
\tilde{F}(k)=\!\int\!F(x)\,{\rm e}^{ik\cdot x}\frac{dx}{(2\pi)^4},
\]
i.e., the factor $1/(2\pi)^4$ attaches to the inverse transform.}
\begin{equation}
\left\langle A_{\mu}^{\ast a}(k)A_{\nu}^b(k^{\prime})\right\rangle=
(A_{\mu}^{\ast a}A_{\nu}^b)_{\omega{\bf k}\,}\delta(k-k^{\prime}).
\label{eq:2u}
\end{equation}
Substituting Eqs.(\ref{eq:2r})\,--\,(\ref{eq:2y}) into (\ref{eq:2e}), taking into account a decomposition
of the gluon propagator in the covariant gauge, Eq.\,(B.1), and the properties of projection operators (B.4),
we find a required form of the spectral density within the limits of the classical model
\begin{equation}
(A_{\mu}^{\ast a}A_{\nu}^b)_{\omega{\bf k}}=
2\,\frac{g^2}{(2\pi)^3}\;\delta^{ab}\!\!\!\!\!\sum\limits_{\;\zeta=\,\textmd{Q},\,\bar{\textmd{Q}},\,\textmd{G}\,}
\!\biggl(\frac{C_2^{(\zeta)}}{d_A}\biggr)
\!\int\!{\bf p}^2 f_{|{\bf p}|}^{(\zeta)}\,\frac{d|{\bf p}|}{2\pi^2}\,
\label{eq:2i}
\end{equation}
\[
\times\Bigl\{
a_tP_{\mu\nu}(k)\,|\!\,^{\ast}\!\Delta^t(k)|^{\,2} + \,a_l\,Q_{\mu\nu}(k)\,|\!\,^{\ast}\!\Delta^l(k)|^{\,2}
\Bigr\}.
\]
\indent
On the other hand, the spectrum of soft gluon modes in an equilibrium QGP can be defined by means of the
fluctuation-dissipation theorem. Here, we can use the well-known expression for the spectral density of fluctuations of an
electromagnetic field \cite{lifshitz_1980, rytov_book_1988} with a minimal extension to the color degrees of freedom
\begin{equation}
(A_{\mu}^{\ast a}A_{\nu}^b)_{\omega{\bf k}}=-\frac{1}{(2\pi)^4}\,i\delta^{ab}
\left(\!\frac{\Theta_B(\omega,T)}{\omega}\right)\!
\Bigl\{{\cal D}^R_{\mu\nu}(k) - ({\cal D}^R_{\nu\mu}(k))^{\ast}\Bigr\},
\label{eq:2o}
\end{equation}
where
\begin{equation}
\Theta_B(\omega,T)=\frac{\hbar\omega}{2}\coth\left(\!\frac{\hbar\omega}{2k_BT}\right)
\label{eq:2p}
\end{equation}
is the mean energy of a quantum bosonic oscillator and $k_B$ is the Boltzmann constant. As for the factor $1/(2\pi)^4$ on the right-hand side
of (\ref{eq:2o}), see footnote \ref{foot_1}. In the semiclassical approximation the gluon (retarded) propagator should be taken
in the HTL-approximation, Eq.\,(B.1). Then in the limit $\hbar\rightarrow 0$, the expression (\ref{eq:2o}) results in
\begin{equation}
(A_{\mu}^{\ast a}A_{\nu}^b)_{\omega{\bf k}}=
\frac{1}{(2\pi)^4}\,\delta^{ab}\!
\left(\!\frac{2k_BT}{\omega}\right)\!
{\rm Im}(\!\,^{\ast}{\cal D}^R_{\mu\nu}(k))
\label{eq:2a}
\end{equation}
\[
=\frac{1}{(2\pi)^4}\,\delta^{ab}\!
\left(\!\frac{2k_BT}{\omega}\right)\!
\Bigl\{P_{\mu\nu}(k)\,{\rm Im}(\!\,^{\ast}\!\Delta^{-1\,t}(k))^{\ast}\,|\!\,^{\ast}\!\Delta^{t}(k)|^{\,2} +
Q_{\mu\nu}(k)\,{\rm Im}(\!\,^{\ast}\!\Delta^{-1\,l}(k))^{\ast}\,|\!\,^{\ast}\!\Delta^{l}(k)|^{\,2}
\Bigr\}.
\]
In the rightmost-hand side of the previous equation we have taken into account the obvious identity
\[
{\rm Im}\,^{\ast}\!\Delta^{t,\,l}(k)=
{\rm Im}(\!\,^{\ast}\!\Delta^{-1\,t,\,l}(k))^{\ast}\,|\!\,^{\ast}\!\Delta^{t,\,l}(k)|^{\,2}.
\]
The requirement of consistency of (\ref{eq:2i}) with expression (\ref{eq:2a}) leads to the conditions
\begin{equation}
2\,\frac{g^2}{(2\pi)^3}\Biggl[\sum\limits_{\;\zeta=\,\textmd{Q},\,\bar{\textmd{Q}},\,\textmd{G}\,}
\!\biggl(\frac{C_2^{(\zeta)}}{d_A}\biggr)
\!\int\!{\bf p}^2 f_{|{\bf p}|}^{(\zeta)}\,\frac{d|{\bf p}|}{2\pi^2}\Biggr]a_{t,\,l}=
\frac{1}{(2\pi)^4}\left(\!\frac{2k_BT}{\omega}\right)
{\rm Im}(\!\,^{\ast}\!\Delta^{-1\,t,\,l}(k))^{\ast}.
\label{eq:2s}
\end{equation}
For definiteness we will consider the condition of consistency for the longitudinal part of the expressions (\ref{eq:2i}) and (\ref{eq:2a}). By virtue of
(B.1) and (B.2), we have
\[
{\rm Im}(\!\,^{\ast}\!\Delta^{-1\,l}(\omega, {\bf k}))^{\ast}=-k^2\frac{m_D^2}{2|{\bf k}|^3}\;\omega\pi
\,\theta({\bf k}^2-\omega^2),
\]
where $m_D^2$ is the Debye screening mass.

Furthermore, in the covariant gauge the longitudinal projector $Q^{\mu\nu}(q)$ in the rest frame of the heat bath, is equal to
\[
Q^{\mu\nu}(k)=
-\,\frac{1}{k^2}\!\left(
\begin{array}{cc}
{\bf k}^2 & -\omega{\bf k}\\
-\omega{\bf k} &  \omega^2\,
\displaystyle\frac{{\bf k}\otimes{\bf k}}{{\bf k}^2}
\end{array}
\right)
\]
and therefore, $a_l$ is defined as follows:
\[
a_l\equiv Q^{\mu\nu}(k)\!\!
\int\!\frac{d\Omega_{{\bf v}}}{4\pi}\;v_{\mu}v_{\nu}\,\delta(v\cdot k)=
-\frac{k^2}{|{\bf k}|^3}\,\theta({\bf k}^2-\omega^2).
\]
Here, we have used the integration formulae over a solid angle (B.10). Based on the above-mentioned,
the condition of consistency (\ref{eq:2s}) will be identically fulfilled for the longitudinal part (and as it is easy to see,
it is also valid for the transverse part), if the Debye screening mass is defined by a relation
\begin{equation}
m_D^2=\frac{2\hspace{0.03cm}g^2\!\!\!\sum\limits_{\;\zeta=\,\textmd{Q},\,\bar{\textmd{Q}},\,\textmd{G}\,}
\!\biggl(\displaystyle\frac{C_2^{(\zeta)}}{d_A}\biggr)\!
\!\int\!{\bf p}^2 f_{|{\bf p}|}^{(\zeta)}\,\frac{d|{\bf p}|}{2\pi^2}}
{k_BT}.
\label{eq:2d}
\end{equation}
\indent
It should be noted, however, that the expression on the right-hand side of (\ref{eq:2d}) represents a definition for the {\it classical\,} Debye mass squared, while
on the left-hand side $m^2_D$ stands for the {\it quantum} Debye mass squared. Let us recall that the latter appears in the definition of the HTL-resummed
gluon propagator (B.1)\,--\,(B.3). By virtue of the fact that the classical Debye mass differs from the quantum one, this circumstance results in violation of the FDT
under this `naive' approach. This fact was first noted in the works \cite{litim_1999}. The matter is that in the average (\ref{eq:2e}) we use the procedure of
the classical statistical average, usually accepted in a high-temperature Abelian plasma. However, for a hot non-Abelian plasma it is not quite true. To derive the correct
quantum value for the Debye mass (and thereby to recover the FDT) following \cite{litim_1999}, we should replace the classical average in (\ref{eq:2e}) by
the quantum one that holds for the systems close to thermal equilibrium. In this case it reduces to the following replacements in (\ref{eq:2e})
\begin{equation}
\int\!{\bf p}^2 f_{|{\bf p}|}^{(\zeta)}\,\frac{d|{\bf p}|}{2\pi^2}
\,\Rightarrow\,
\left\{
\begin{array}{ll}
\displaystyle\int\!{\bf p}^2 f_{|{\bf p}|}^{(\textmd{G})}\!\left(1 + f_{|{\bf p}|}^{(\textmd{G})}\right)\displaystyle\frac{d|{\bf p}|}{2\pi^2}, & \mbox{for } \zeta=\textmd{G}, \\
N_f\!\!\displaystyle\int\!{\bf p}^2 f_{|{\bf p}|}^{(\textmd{Q},\,\bar{\textmd{Q}})}\!\left(1 - f_{|{\bf p}|}^{(\textmd{Q},\,\bar{\textmd{Q}})}\right)\displaystyle\frac{d|{\bf p}|}{2\pi^2},
& \mbox{for } \zeta=\textmd{Q},\,\bar{\textmd{Q}},
\end{array}
\right.
\label{eq:2f}
\end{equation}
where $N_f$ is the number of active (massless) quark flavors. Then taking into account the last circumstance we will have the following expression for the Debye
mass instead of (\ref{eq:2d})
\begin{equation}
\begin{split}
m_D^2=
\biggl(\frac{2\hspace{0.03cm}g^2}{k_B T}\biggr)\!\!
\int\!{\bf p}^2\biggl\{\!&\biggl(\frac{C_2^{(\textmd{G})}}{d_A}\biggr) f_{|{\bf p}|}^{(\textmd{G})}\left(1 + f_{|{\bf p}|}^{(\textmd{G})}\right)+\\
+\,
\biggl[N_f&\biggl(\frac{C_2^{(\textmd{Q})}}{d_A}\biggr)f_{|{\bf p}|}^{(\textmd{Q})}\left(1 - f_{|{\bf p}|}^{(\textmd{Q})}\right) + (\textmd{Q}\rightarrow\bar{\textmd{Q}})\biggr]\biggr\}
\frac{d|{\bf p}|}{2\pi^2}.
\end{split}
\label{eq:2g}
\end{equation}
For values of the quadratic Casimirs $C_2^{(\zeta)}$ specified in Appendix A, the expression (\ref{eq:2g}) correctly reproduces the quantum Debye mass squared as it was defined in the 
high-temperature QCD \cite{kalashnikov_1999}.\\
\indent
Thus the simple model for the classical description of the hot non-Abelian plasma with the known proviso enables us to exactly reproduce the FDT relation for soft bosonic fluctuations.

\section{\bf Correlation function of soft Fermi excitations}
\setcounter{equation}{0}

We proceed now to the problem of calculation of the correlation function for thermal
fermionic excitations. Within the model suggested in \cite{markov_NPA_2007,markov_IJMPA_2010}
for a description of the soft and hard fermion degrees of freedom of the system under consideration,
the soft spinor fields $\psi$ and $\bar{\psi}$ induced by a free spin-1/2 hard test particle, are
\begin{equation}
\begin{split}
& \psi_{\alpha}^i(q)=
-\,^{\ast}\!S_{\alpha\beta}(q)\,
\eta_{\beta}^i(q;{\bf x}_{0}),\quad q=(\omega,{\bf q}),\\
& \bar{\psi}_{\alpha}^i(-q)=
\bar{\eta}^i_{\beta}(-q;{\bf x}_0)\,^{\ast}\!S_{\beta\alpha}(-q),
\end{split}
\label{eq:3q}
\end{equation}
where a color source $\eta_{\beta}^i$ of the particle has the form
\begin{equation}
\eta_{\beta}^i(q;{\bf x}_{0})=
\frac{\,g}{(2\pi)^3}\,\theta^i\chi_{\beta}\,
\delta(v\cdot q)\,
{\rm e}^{-i{\bf q}\cdot\,{\bf x}_{0}},
\label{eq:3w}
\end{equation}
and $\bar{\eta}^i(-q;{\bf x}_0)=\eta^{\dagger{i}}(q;{\bf x}_{0})\gamma^0$. Here, $\theta^i$ is a
Grassmann color charge and $\chi_{\beta}$ is a $c$-number spinor describing a polarization state of the
spinning particle.\\
\indent
Let us consider the correlation function
$\left\langle\psi_{\alpha}^i(q)\bar{\psi}_{\beta}^j(-q^{\prime})\right\rangle$. First of all we will discuss
a question what is meant by the angular brackets of averaging. As in the bosonic case
the average includes the integration over every possible initial positions ${\bf x}_0$ of the hard test parton.
Besides, the average also includes the integration over Grassmann color charges with the measure $d\theta d\theta^{\dagger}$
specified in Appendix C, Eqs.\,(C.2) and (C.5). Finally, it is necessary to define the correct statistical factor.
As was shown in \cite{markov_NPA_2007}, the relevant statistical factor here must be
\begin{equation}
\sum\limits_{\;\zeta=\textmd{Q},\,\bar{\textmd{Q}}\,}
\!\int\!{\bf p}^2
\left[\,f_{\bf p}^{(\zeta)} + f_{\bf p}^{(\textmd{G})}\,\right]
\frac{d{\bf p}}{(2\pi)^3}.
\label{eq:3ww}
\end{equation}
\indent
From the above-mentioned and the formula of integration with respect to Grassmann color charges (C.6),
we obtain
\begin{equation}
\begin{split}
\bigl\langle\psi_{\alpha}^i(q)&
\bar{\psi}_{\beta}^j(-q^{\prime})\bigr\rangle=
-\frac{\,g^2}{(2\pi)^6}\;\delta^{ij}\!\!\!
\sum\limits_{\;\zeta=\,\textmd{Q},\,\bar{\textmd{Q}}\,}\!\!\!\left(\frac{C_{\theta}^{(\zeta)}}{N_c}\right)
\!\int\!{\bf p}^2\!
\left[\,f_{|{\bf p}|}^{(\zeta)} + f_{|{\bf p}|}^{(\textmd{G})}\,\right]\!
\frac{d|\,{\bf p}|}{2\pi^2}\\
&\times\!\!\int\!\frac{d\Omega_{{\bf v}}}{4\pi}
\left(\!\,^{\ast}\!S(q)\chi\right)_{\alpha}
\left(\bar{\chi}\,^{\ast}\!S(-q^{\prime})\right)_{\beta}
\delta(\omega-{\bf v}\cdot{\bf q})\,\delta(q-q^{\prime}).
\end{split}
\label{eq:3e}
\end{equation}
We notice that unlike the bosonic case now it is impossible to write down the integral over the magnitude of hard momentum $|{\bf p}|$
in the form of an independent multiplier since the spinors $\chi$ and $\bar{\chi}$ in the second line of the above expression
also implicitly depend on $|{\bf p}|$, see below.\\
\indent
Let us analyze the integrand in (\ref{eq:3e}) with the quark propagators. We shall restrict ourselves to the case of fully unpolarized
state of (massless) hard test particle. As was shown in Appendix C of the paper \cite{markov_NPA_2007}, the following replacement
\[
\chi_{\alpha}\bar{\chi}_{\beta}\rightarrow\frac{1}{4|{\bf p}|}\,(v\cdot\gamma)_{\alpha\beta}
\]
is valid. Then we get
\begin{equation}
\left(\!\,^{\ast}\!S(q)\chi\right)_{\alpha}\left(\bar{\chi}\,^{\ast}\!S(-q)\right)_{\beta}=
\frac{1}{4|{\bf p}|}\,\left(\!\,^{\ast}\!S(q)(v\cdot\gamma)\,^{\ast}\!S(-q)\right)_{\alpha\beta}.
\label{eq:3r}
\end{equation}
An explicit form of the quark propagator in the HTL-approximation is given in Appendix B, Eq.\,(B.5).
Within this approximation the propagator represents an expansion in terms of the spinor projectors $h_{\pm}(\hat{\bf q})$, whose
explicit forms are defined by equation (B.6).
Our first task is to present the matrix $(v\cdot\gamma)\equiv\gamma^0-{\bf v}\cdot \boldsymbol{\gamma}$ entering into the right-hand side of
(\ref{eq:3r}) in the form of a similar expansion. For the $\gamma^0$ matrix we have
\[
\gamma^0=h_{+}(\hat{\bf q})+h_{-}(\hat{\bf q}).
\]
Then we write a scalar product ${\bf v}\cdot\boldsymbol{\gamma}=v^i\delta^{ij}\gamma^j$ in the form of an expansion in 
three-dimensional transverse and longitudinal projectors with respect to the vector of momentum ${\bf q}$
\[
v^i(\delta^{ij}-\hat{q}^i\hat{q}^j)\gamma^j +
({\bf v}\cdot\hat{\bf q})(\boldsymbol{\gamma}\cdot\hat{\bf q}),\quad
\hat{\bf q} \equiv \frac{\bf q}{|{\bf q}|}.
\]
Here, in the second term we can also write
\[
(\boldsymbol{\gamma}\cdot\hat{\bf q}) = -(h_{+}(\hat{\bf q})-h_{-}(\hat{\bf q})).
\]
From what has been said, we finally derive
\[
(v\cdot\gamma)= (1+{\bf v}\cdot\hat{\bf q})h_{+}(\hat{\bf q})+
(1-{\bf v}\cdot\hat{\bf q})h_{-}(\hat{\bf q})-
\boldsymbol{\gamma}\cdot(\hat{\bf q}\times({\bf v}\times\hat{\bf q})).
\]
The last term on the right-hand side of the above expression upon integrating over a solid angle $d\Omega_{\bf v}/4\pi$ in view of the function
$\delta(\omega-{\bf v}\cdot\hat{\bf q})$, will turn into zero. Therefore, subsequently this term will be omitted.
Substituting the expression for $(v\cdot\gamma)$ obtained by this means into (\ref{eq:3r}), taking into account the structure of the quark
propagator and the properties of the $h_{\pm}(\hat{\bf q})$ matrices (B.9), we derive instead of
(\ref{eq:3r})
\[
\bigl(\!\,^{\ast}\!S(q)\chi\bigr)\!\otimes\!
\bigl(\bar{\chi}\,^{\ast}\!S(-q)\bigr)=
\]
\[
=- \frac{1}{4|{\bf p}|}\,
\Bigl[\bigl(1-{\bf v}\cdot\hat{\bf q}\bigr)h_{+}(\hat{\bf q})\,|\!\,^{\ast}\!\triangle_{+}(q)|^2
+\,\bigl(1+{\bf v}\cdot\hat{\bf q}\bigr)h_{-}(\hat{\bf q})\,|\!\,^{\ast}\!\triangle_{-}(q)|^2
\Bigr].
\]
\indent
The angular integration $d\Omega_{\bf v}$ in (\ref{eq:3e}) is easily performed employing Eq.\,(B.10). We present the final expression for
the correlator (\ref{eq:3e}) in the form similar to (\ref{eq:2u}), i.e.,
\[
\left\langle\psi_{\alpha}^i(q)\bar{\psi}_{\beta}^j(-q^{\prime})\right\rangle=
(\psi_{\alpha}^i\bar{\psi}_{\beta}^j)_{\omega{\bf q}}\,\delta(q-q^{\prime}),
\]
where
\begin{equation}
(\psi_{\alpha}^i\bar{\psi}_{\beta}^j)_{\omega{\bf q}}=
\delta^{ij\,}\frac{1}{2(2\pi)^3}\,
\frac{\omega_0^2}{2|{\bf q}|}
\label{eq:3t}
\end{equation}
\[
\times\!\left[\left(1-\frac{\omega}{|{\bf q}|}\right)\!(h_{+}(\hat{\bf q}))_{\alpha\beta}|\!\,^{\ast}\!\triangle_{+}(q)|^2
\,+\,\left(1+\frac{\omega}{|{\bf q}|}\right)\!(h_{-}(\hat{\bf q}))_{\alpha\beta}|\!\,^{\ast}\!\triangle_{-}(q)|^2
\right]\!\theta({\bf q}^2-\omega^2)
\]
is the spectral density for soft quark fluctuations, and
\begin{equation}
\omega_0^2=-\frac{g^2}{4\pi^2}\!\!
\sum\limits_{\;\zeta=\,\textmd{Q},\,\bar{\textmd{Q}}\,}\!\left(\frac{C_{\theta}^{(\zeta)}}{N_c}\right)\!
\!\int\!|{\bf p}|
\left[\,f_{|{\bf p}|}^{(\zeta)} + f_{|{\bf p}|}^{(\textmd{G})}\,\right]\!d|\,{\bf p}|.
\label{eq:3y}
\end{equation}
From the last expression we see that the quantity $\omega_0^2$ in exact coincides with the fermion plasma frequency squared if we set
\begin{equation}
C_{\theta}^{(\textmd{Q})}\! = C_{\theta}^{(\bar{\textmd{Q}})} \equiv -\,C_FN_c.
\label{eq:3u}
\end{equation}
Recall that the constant $C_{\theta}^{(\zeta)}$ is connected with an integral of motion for a system of dynamical equations
(C.1) and it is the only free parameter in the model under consideration. The above-mentioned requirement uniquely fixes this constant. It is interesting to note
that with the choice of (\ref{eq:3u}) the constant $C_{\theta}^{(\zeta)}$ correct to a sign, coincides with the quark quadratic Casimirs (A.3) by virtue
of the identity
\[
C_F N_c = T_{F\,} d_A.
\]
Thus this choice looks quite reasonable.\\
\indent
In addition, one further remark is in order. In contrast to the expression for the Debye screening mass (\ref{eq:2d}), we
have immediately derived the expression for the fermion plasma frequency\footnote{\,It is pertinent at this point to note that, generally
speaking, analog of quantity $\omega_0$ is not available in a classical plasma, in contrast to the Debye mass $m_D$.} (\ref{eq:3y}),
as it appears in calculating within the high-temperature QCD \cite{klimov_1981, weldon_1982}. This circumstance suggests that
(\ref{eq:3ww}) represents not classical, but quantum average. Strictly speaking, the statistical average should be written in the following form:
\begin{equation}
\sum\limits_{\;\zeta=\,\textmd{Q},\,\bar{\textmd{Q}}\;}
\!\int\!{\bf p}^2
\Bigl[f_{\bf p}^{(\textmd{G})}\Bigl(1 - f_{\bf p}^{(\zeta)}\Bigr) + f_{\bf p}^{(\zeta)}\Bigl(1 + f_{\bf p}^{(\textmd{G})}\Bigr)\Bigr]
\frac{d{\bf p}}{(2\pi)^3},
\label{eq:3i}
\end{equation}
which is `fermionic' analog of the quantum average (\ref{eq:2f}). A somewhat unusual mixed form of arrangement of the distribution
functions for thermal particles is caused by the following fact: in the emission (absorption) process of soft Fermi-excitations by a hard particle,
this particle changes own statistics (unlike the emission or absorption processes of soft Bose-excitations, as it is presented in the quantum average (\ref{eq:2f})).
Nevertheless, from the rigorous definition for the statistical average (\ref{eq:3i}) we see that all nonlinear terms in the distribution functions, are canceled
in exact, and thereby we lead to the simpler expression (\ref{eq:3ww}). It is curious to note that quantum average (\ref{eq:3i}) formally coincides
with classical one, which follows from (\ref{eq:3i}) in the limit $f_{\vert {\bf p} \vert}^{(G)}$, $f_{\vert {\bf p} \vert}^{(Q,\,\bar{Q})}\!\ll 1$.\\
\indent
By the next step expression (\ref{eq:3t}) should be compared with the spectral density that follows from the fluctuation-dissipation
theorem for soft fermionic fluctuations. Unfortunately, as opposed to the bosonic case we could not find somewhere in the literature the FDT relation
we need. Therefore, two forthcoming sections will be devoted to derivation of the required relation.

\section{\bf Fluctuation-dissipation theorem for source $\eta_{\alpha}^i$}
\setcounter{equation}{0}

In spite of the fact that we are properly interested in the FDT for soft fermionic fluctuations, conceptually this theorem can be
defined to a limited extent, more easily for fluctuations of the hard Grassmann-valued source $\eta_{\alpha}^i$. The fluctuations of
sources $\eta_{\alpha}^i$ and $\bar{\eta}_{\alpha}^i$ are thermodynami\-cal\-ly mutually complementary quantities with the
fluctuations of soft fermionic fields $\bar{\psi}_{\alpha}^i$ and $\psi_{\alpha}^i$. Therefore, knowing the FDT for the first
quantities, it is not difficult to restore the FDT for the second ones.\\
\indent
The most rigorous derivation of the fluctuation-dissipation relations is based on the technique of the
{\it spectral representation} for correlation functions \cite{fradkin_1965, heinz_1986}.
In the interests of brevity we have chosen here a more simple way without using such a representation
(see, for example \cite{zubarev_book_1974, akhiezer_book_1986}). First of all we present the total
fermion field in the system under study as a sum of two parts
\begin{equation}
\psi_{\alpha}^i + \Psi_{\alpha}^i,
\label{eq:4q}
\end{equation}
where the spinors $\psi_{\alpha}^i$ and $\Psi_{\alpha}^i$ contain the soft and hard momenta, respectively.\\
\indent
Let us assume that our many-body system is described by the following Hamiltonian $\hat{H}(t)=\hat{H}_0+\hat{H}_t^1$ with
the interaction term (in the Schr\"odinger picture)
\begin{equation}
\hat{H}_t^1=\!\int\!d{\bf x}\Bigl[\,\hat{\bar{\eta}}^i_{\alpha}({\bf x})\psi^i_{\alpha,\,{\rm ext}}({\bf x},t)
+\bar{\psi}^i_{\alpha,\,{\rm ext}}({\bf x},t)\hat{\eta}^i_{\alpha}({\bf x})\Bigr],
\label{eq:4w}
\end{equation}
where $\psi^i_{\alpha,\,{\rm ext}}({\bf x},t)$ and $\bar{\psi}^i_{\alpha,\,{\rm ext}}({\bf x},t)$ are
``small'' time-dependent Grassmann-valued external fermion fields. The color sources $\hat{\eta}^i_{\alpha}({\bf x})$
and $\hat{\bar{\eta}}^i_{\alpha}({\bf x})$ are expressed in a certain way through the hard part $\hat{\Psi}_{\alpha}^i$
(and $\hat{\bar{\Psi}}_{\alpha}^i$) of decomposition (\ref{eq:4q}), i.e., in other words, through the wave functions of hard
half-spin particles of the system. The hat above points to the operator nature of the quantity under consideration. In what follows,
as a simplification in order not to overburden the subsequent formulae, the hat will be systematically omitted. As usual, we suppose that
the external perturbation is switched off at $t=-\infty$, i.e.
\[
H_t^1\Bigr|_{t=-\infty}=0.
\]
\indent
Furthermore, we introduce into consideration a statistical operator $\rho$. This operator obeys the quantum Liouville equation
\begin{equation}
i\hbar\,\frac{\partial\rho}{\partial t}=
[H_0+H_t^1,\rho].
\label{eq:4e}
\end{equation}
Here, the bracket symbol signifies the commutator. As the initial condition we make use the Gibbs grand canonical ensemble
\[
\rho\bigr|_{t=-\infty} \equiv \rho_0=\exp\{\Omega-\beta(H_0-\mu N)\},
\]
where $\Omega$ is the thermodynamical grand potential, $\beta=1/k_B T$, and $\mu$ is a chemical potential associated with
a baryon number
\begin{equation}
N=\!\int\!\bar{\Psi}^i({\bf x})\gamma_0{\Psi}^i({\bf x})\,d{\bf x}.
\label{eq:4r}
\end {equation}
\indent
Equation (\ref{eq:4e}) may be solved iteratively, and in a linear approximation in $H_t^1$ we have
\begin{equation}
\rho(t)\simeq\rho_0+
\frac{1}{i\hbar}\!\int\limits_{-\infty}^{t}\!\left[H_{t^{\prime}}^1(t^{\prime}-t),\rho_0\right]\!dt^{\prime},
\label{eq:4t}
\end{equation}
where $H_{t^{\prime}}^1(t^{\prime}-t)={\rm e}^{iH_0(t^{\prime}-t)/\hbar}H_{t^{\prime}}^1\,{\rm e}^{-iH_0(t^{\prime}-t)/\hbar}$.
The expectation value of the operator $\eta_{\alpha}^i({\bf x})$ is defined by the usual formula
\[
\langle\eta_{\alpha}^i\rangle=
{\rm Sp}(\rho\eta_{\alpha}^i).
\]
Putting the approximation (\ref{eq:4t}) into the last expression and making use the cyclic invariance of the trace, we have
\[
\langle\eta_{\alpha}^i\rangle=\langle\eta_{\alpha}^i\rangle_{0}+
\frac{1}{i\hbar}\!\int\limits_{-\infty}^{+\infty}\!\!\theta(t-t^{\prime})
\,{\rm Sp}\!\left(\rho_0\!\left[\eta_{\alpha}^i({\bf x},t),H_{t^{\prime}}^1(t^{\prime})\right]\right)\!dt^{\prime}.
\]
Here, $\eta_{\alpha}^i({\bf x},t)={\rm e}^{iH_0 t/\hbar}\eta_{\alpha}^i({\bf x})\,{\rm e}^{-iH_0 t/\hbar}$ is the $\eta_{\alpha}^i$ operator
in the Heisenberg picture with respect to the Hamiltonian $\hat{H}_0$ and $\langle\ldots\rangle_0\equiv{\rm Sp}(\rho_0\dots)$ is an averaging over the Gibbs distribution of a global
thermal equilibrium configuration. By virtue of local color neutrality and zero fermionic number,
$\langle\eta_{\alpha}^i\rangle_{0}=0$.\\
\indent
Further, substituting the external perturbation (\ref{eq:4w}) into the preceding equation, we find
\[
\langle\eta_{\alpha}^i\rangle=\!\int\!G^{(R)ij}_{\alpha\beta}(x-x^{\prime})
\psi^j_{\beta,\,{\rm ext}}(x^{\prime})\,dx^{\prime}\,+
\int\!\bar{\psi}^j_{\beta,\,{\rm ext}}(x^{\prime})
F^{ji}_{\beta\alpha}(x^{\prime}-x)\,dx^{\prime},
\quad x=(t,{\bf x}),
\]
where we have introduced the double-time Green's functions
\[
G^{(R)ij}_{\alpha\beta}(x-x^{\prime})=\frac{1}{i\hbar}\,\theta(t-t^{\prime})\,
{\rm Sp}\!\left(\rho_0\!\left\{\eta_{\alpha}^i({\bf x},t),\bar{\eta}_{\beta}^j({\bf x}^{\prime},t^{\prime})\right\}\right),
\]
\[
F^{ji}_{\beta\alpha}(x^{\prime}-x)=-\frac{1}{i\hbar}\,\theta(t-t^{\prime})\,
{\rm Sp}\!\left(\rho_0\!\left\{\eta_{\alpha}^i({\bf x},t),\eta_{\beta}^j({\bf x}^{\prime},t^{\prime})\right\}\right).
\]
Here $\{\,,\}$ denotes the anticommutator. The first function represents the usual retarded propagator for the $\eta$ field,
the second one is the anomalous propagator. The anomalous Green's function in the situation under consideration equals zero.
In addition to the retarded Green's function we also introduce the double-time advanced Green's function
\[
\hspace{0.3cm}
G^{(A)ij}_{\alpha\beta}(x-x^{\prime})=-\frac{1}{i\hbar}\,\theta(t^{\prime}-t)\,
{\rm Sp}\!\left(\rho_0\!\left\{\eta_{\alpha}^i({\bf x},t),\bar{\eta}_{\beta}^j({\bf x}^{\prime},t^{\prime})\right\}\right).
\]
\indent
To obtain the required fluctuation-dissipation theorem following A.I.\,\,Akhiezer and S.V. Peletminski \cite{akhiezer_book_1986}, we introduce
into consideration a new pair of the Green's functions more simple in a structure:
\begin{equation}
{\cal J}^{ji}_{\beta\alpha}(x-x^{\prime})=
\langle\bar{\eta}_{\beta}^j({\bf x}^{\prime},t^{\prime})\,\eta_{\alpha}^i({\bf x},t)\rangle_0
\label{eq:4y}
\end{equation}
and
\begin{equation}
\hspace{0.1cm}
{\cal J}^{ij}_{\alpha\beta}(x^{\prime}-x)=
\langle\eta_{\alpha}^i({\bf x},t)\,\bar{\eta}_{\beta}^j({\bf x}^{\prime},t^{\prime})\rangle_0.
\label{eq:4u}
\end{equation}
Let us define a relation between these two functions. For (\ref{eq:4y}) we have
\[
{\cal J}^{ji}_{\beta\alpha}(x-x^{\prime})=
{\rm Sp}\,\rho_0\!\left(\bar{\eta}_{\beta}^j(x^{\prime})\,\eta_{\alpha}^i(x)\right)=
{\rm Sp}\,\rho_0\!\left[\left(\rho_0^{-1}\eta_{\alpha}^i(x)\rho_0\right)\bar{\eta}_{\beta}^j(x^{\prime})\right],
\]
where in turn
\[
\rho_0^{-1}\eta_{\alpha}^i(x)\rho_0=
{\rm e}^{\,\beta(H_0-\mu N)}\eta_{\alpha}^i({\bf x},t)\,{\rm e}^{-\beta(H_0-\mu N)}=
{\rm e}^{-\beta\mu N}\eta_{\alpha}^i({\bf x},t-i\hbar\beta)\,{\rm e}^{\,\beta\mu N}.
\]
The source $\eta_{\alpha}^i$ is generally an odd function of the operator $\Psi_{\alpha}^i$ and an even one of the operator $\bar{\Psi}_{\alpha}^i$.
In our most simple situation, the $\eta_{\alpha}^i$ linearly depends on $\Psi_{\alpha}^i$, and it does not depend on $\bar{\Psi}_{\alpha}^i$ at all.
By using the definition of baryon number operator (\ref{eq:4r}) and the anticommutation relations for the $\Psi$-operators, we
have
\[
{\rm e}^{-\beta\mu N}\eta_{\alpha}^i({\bf x},t-i\hbar\beta)\,{\rm e}^{\,\beta\mu N}=
\eta_{\alpha}^i({\bf x},t-i\hbar\beta)\,{\rm e}^{\,\beta\mu},
\]
that immediately results in the following relation between (\ref{eq:4y}) and (\ref{eq:4u})
\[
{\cal J}^{ij}_{\alpha\beta}({\bf x}^{\prime}-{\bf x},t^{\prime}-t)=
{\cal J}^{ji}_{\beta\alpha}({\bf x}-{\bf x}^{\prime},t-t^{\prime}+i\hbar\beta)
\,{\rm e}^{-\beta\mu},
\]
or in terms of Fourier components it reads
\begin{equation}
{\cal J}^{ij}_{\alpha\beta}(-{\bf q},-\omega)=
{\cal J}^{ji}_{\beta\alpha}({\bf q},\omega)
\,{\rm e}^{\,\beta(\hbar\omega-\mu)}.
\label{eq:4i}
\end{equation}
The fact of appearance of the chemical potential in an explicit form on the right-hand side of (\ref{eq:4i}) is rather important in subsequent
discussion\footnote{\,In a number of monographes devoted to quantum statistical mechanics (see, for example, \cite{bogolubov_1982})
the generalized Hamiltonian $H_0^{\prime}\equiv H_0-\mu N$ is introduced instead of the Hamiltonian $H_0$. This leads to the fact that
on the right-hand side of the quantum Liouville equation (\ref{eq:4e}), additional term proportional to $\mu$, appears, i.e. instead
of (\ref{eq:4e}) now we will have
\[
i\hbar\,\frac{\partial\rho}{\partial t}=
[H_0+H_t^1,\rho]-\mu\,[N,\rho].
\]
Repeating the reasoning resulting in (\ref{eq:4i}), it is not difficult to see that instead of (\ref{eq:4i}) in
this case we will have ${\cal J}^{ij}_{\alpha\beta}(-{\bf q},-\omega)=
{\cal J}^{ji}_{\beta\alpha}({\bf q},\omega)\,{\rm e}^{\,\beta\hbar\omega}$, i.e., the chemical potential has disappeared.
It will be entangled in the definitions of the correlators (\ref{eq:4y}) and (\ref{eq:4u}) in a complicated implicit way.
We notice that practically the same problem of two possible ways of constructing the fermion propagator at finite temperature and density
was discussed by A. Ni\'egawa in \cite{niegawa_2002}.}.\\
\indent
The retarded and advanced Green's functions are expressed in terms of the correlation functions (\ref{eq:4y}) and (\ref{eq:4u})
as follows:
\[
G^{(R,\,A)\,ij}_{\alpha\beta}(x-x^{\prime})=\pm\,\frac{1}{i\hbar}\,\theta(\pm(t-t^{\prime}))
\!\left[{\cal J}^{ij}_{\alpha\beta}(x^{\prime}-x)+{\cal J}^{ji}_{\beta\alpha}(x-x^{\prime})\right].
\]
Further, in terms of Fourier components the above expression, in view of Eq.\,(\ref{eq:4i}), can be written as
\[
G^{(R,\,A)\,ij}_{\alpha\beta}({\bf q},\omega)=\frac{1}{2\pi\hbar}\!
\int\limits_{-\infty}^{+\infty}\!d\omega^{\prime}\,
\frac{{\cal J}^{ji}_{\beta\alpha}({\bf q},\omega^{\prime})}{\omega-\omega^{\prime}\pm i0}\,
\bigl({\rm e}^{\beta(\hbar\omega^{\prime}-\mu)}+1\bigr),
\]
whence, in particular, it follows that
\begin{equation}
G^{(R)\,ij}_{\alpha\beta}({\bf q},\omega)-G^{(A)\,ij}_{\alpha\beta}({\bf q},\omega)=
\frac{1}{i\hbar}\,\bigl({\rm e}^{\,\beta(\hbar\omega-\mu)}+1\bigr)
{\cal J}^{ji}_{\beta\alpha}({\bf q},\omega).
\label{eq:4o}
\end{equation}
\indent
The final step in a derivation of the FDT is introducing into consideration the correlation function
for fluctuations of the color source $\eta_{\alpha}^i$. As an initial expression we set the following one:
\begin{equation}
\Xi^{ij}_{\alpha\beta}(x-x^{\prime})\equiv
\frac{1}{2}\,\langle[\eta^i_{\alpha}(x),\bar{\eta}^j_{\beta}(x^{\prime})]\rangle_0.
\label{eq:4p}
\end{equation}
In this definition we have considered the fact that $\langle\eta^i_{\alpha}(x)\rangle_0=\langle\bar{\eta}^i_{\alpha}(x)\rangle_0 = 0$.
The commuta\-tor under the average sign is reflecting statistics of the $\eta^i_{\alpha}(x)$ source. We believe that in the
semiclassical limit the operators $\eta_{\alpha}^i$ and $\bar{\eta}_{\alpha}^i$ turn to classical Grassmann-valued functions.
Notice that the function (\ref{eq:4p}) possesses the property evident from its definition
\[
(\Xi^{ij}(x-x^{\prime}))^{\dagger}=\gamma^{0\,}\Xi^{ji}(x^{\prime}-x)\gamma^0.
\]
\indent
Furthermore, we present (\ref{eq:4p}) in terms of the correlation functions (\ref{eq:4y}) and (\ref{eq:4u})
\[
\Xi^{ij}_{\alpha\beta}(x-x^{\prime})=\frac{1}{2}
\left\{ {\cal J}^{ij}_{\alpha\beta}(x^{\prime}-x)
-{\cal J}^{ji}_{\beta\alpha}(x-x^{\prime})\right\},
\]
or by using its Fourier components, considering (\ref{eq:4i})
\[
\Xi^{ij}_{\alpha\beta}({\bf q},\omega)=
\frac{1}{2}\,\bigl({\rm e}^{\beta(\hbar\omega-\mu)}-1\bigr){\cal J}^{ji}_{\beta\alpha}({\bf q},\omega).
\]
After substituting ${\cal J}^{ji}_{\beta\alpha}({\bf q},\omega)$ from the above equation into (\ref{eq:4o}), we
arrive at the desired fluctuation-dissipation relation
\begin{equation}
\Xi^{ij}_{\alpha\beta}({\bf q},\omega)=\frac{1}{2}\,i\hbar\,
\tanh\frac{1}{2}\beta(\hbar\omega-\mu)
\left\{G^{(R)\,ij}_{\alpha\beta}({\bf q},\omega)-G^{(A)\,ij}_{\alpha\beta}({\bf q},\omega)\!\right\}.
\label{eq:4a}
\end{equation}
We can present the given relation similar to bosonic case (\ref{eq:2o}). Indeed, the partition function
for bosonic oscillator has the form
\[
Z_B=\sum\limits_{n=0}^{\infty}\exp\Bigl\{-\beta\hbar\omega\Bigl(n+\frac{1}{2}\Bigr)\Bigr\}=
\frac{1}{2}\,\frac{1}{\sinh(\beta\hbar\omega/2)},
\]
and thus its mean energy is
\[
\Theta_B(\omega,T)=-\frac{\partial}{\partial\beta}\ln Z_B=
\frac{1}{2}\,\hbar\omega\,\coth\!\left(\frac{\hbar\omega}{2k_BT}\right),
\]
as it was defined in (\ref{eq:2p}). On the other hand, in the case of fermionic oscillator, at finite chemical potential,
the partition function has the form
\[
Z_F=\sum\limits_{n=0,1}\exp\Bigl\{-\beta(\hbar\omega-\mu)\Bigl(n-\frac{1}{2}\Bigr)\Bigr\}=
2\cosh(\beta(\hbar\omega-\mu)/2)
\]
and its mean energy reads
\[
\Theta_F(\omega,\mu,T)=-\frac{\partial}{\partial\beta}\ln Z_F=
-\frac{1}{2}\,(\hbar\,\omega-\mu)\,\tanh\!\left(\frac{\hbar\omega-\mu}{2k_BT}\right).
\]
Taking into account the preceding expression, the FDT (\ref{eq:4a}) can be rewritten as follows:
\[
\Xi^{ij}_{\alpha\beta}({\bf q},\omega)=-i\hbar\,
\frac{\Theta_F(\omega,\mu,T)}{(\hbar\,\omega-\mu)}
\left\{G^{(R)\,ij}_{\alpha\beta}({\bf q},\omega)-G^{(A)\,ij}_{\alpha\beta}({\bf q},\omega)\right\}.
\]

\section{\bf FDT for soft fermionic fluctuations}
\setcounter{equation}{0}

After providing a detailed derivation of the FDT for fluctuations of hard source, we move on to one of the main goals of this article,
namely, to the formulation of the FDT for soft fermionic fluctuations. We can obtain the required FDT from the previously derived
relation (\ref{eq:4a}) by using the fact of mutuality between the fluctuations of generalized forces (in this case these are the sources
$\eta_{\alpha}^i$ and $\bar{\eta}_{\alpha}^i$) and generalized stationary and homogeneous fields in a system (in this case these are the
soft fermionic fields  $\psi_{\alpha}^i$ and $\bar{\psi}_{\alpha}^i$).
This fact was already mentioned at the beginning of Section 4. Now we should understand as the perturbation $H_t^1$ the following expression,
instead of (\ref{eq:4w}):
\[
H_t^1=\!\int\!d{\bf x}\Bigl[\,\bar{\eta}^i_{\alpha,\,{\rm ext}}({\bf x},t)\psi^i_{\alpha}({\bf x})
+\bar{\psi}^i_{\alpha}({\bf x})\eta^i_{\alpha,\,{\rm ext}}({\bf x},t)\Bigr],
\]
where $\bar{\eta}^i_{\alpha,\,{\rm ext}}({\bf x},t)$ and $\eta^i_{\alpha,\,{\rm ext}}({\bf x},t)$ play
a role of the generalized Grassmann-valued external forces that couple to the operators of soft fermionic fields
$\psi^i_{\alpha}({\bf x})$ and $\bar{\psi}^i_{\alpha}({\bf x})$. For these soft fields we introduce the
retarded and advanced Green's functions
\begin{equation}
S^{\,(R,\,A)\,ij}_{\alpha\beta}(x-x^{\prime})=\pm\,\frac{1}{i\hbar}\,\theta(\pm(t-t^{\prime}))\,
{\rm Sp}\!\left(\rho_0\!\left\{\psi_{\alpha}^i({\bf x},t),\bar{\psi}_{\beta}^j({\bf x}^{\prime},t^{\prime})\right\}\right).
\label{eq:5q}
\end{equation}
Here, the average is taken over the Gibbs distribution for a system consisting of the medium and the radiation of soft fermionic
excitations, being at thermal equilibrium with the medium. The $\psi$-operators are understood as averaged over physically infinite\-si\-mal volume
with a given arrangement of all the hard particles in it. Thereby we restrict ourselves to considering only the long-wavelength part of the fermion
radiation. This enables us to express the Green's functions in terms of macroscopical characteristics of the medium: in this case in terms of the
retarded and advanced quark propagators in the HTL-approximation \cite{braaten_1990}.\\
\indent
The system under consideration can be regarded as involving two subsystems, which by the convention can be called the {\it hard} and {\it soft} ones.
As a result of thermal fluctua\-tions there exists the fermion number `pumping over' from one subsystem to another and vice verse. In this regard
the number of hard particles carrying a half-integer spin (more exactly, the difference of the numbers of hard particles and antiparticles) and soft quasiparticles
which also obey Fermi statistics, is not separately conserved but only its sum is. The chemical potentials associated with the hard and soft subsystems
subject to thermal equilibrium, are connected by the relation
\begin{equation}
\mu_{\,\rm hard}=\mu_{\,\rm soft}\equiv\mu.
\label{eq:5w}
\end{equation}
\indent
Let us briefly explain what we understand by mutuality between $\eta_{\alpha}^i$ and $\psi_{\alpha}^i$ \cite{rytov_book_1988}.
Let fluctuations in any continuous system be described by some stationary and homogeneous fields $\xi^{(i)}({\bf x},t),\;
i=1,\ldots,N$ and let $f^{(i)}({\bf x},t)$ be a volume density of the generalized force fields, conjugate in Lagrange's sense
with the generalized coordinates $\xi^{(i)}({\bf x},t)$. The mean power dissipated in the volume $V$ under the action of the
$f^{(i)}({\bf x},t)$ forces, is written as
\begin{equation}
\langle Q\rangle=\int\limits_V
\Bigl\langle f^{(i)}({\bf x},t)\frac{\partial \xi^{(i)}({\bf x},t)}{\partial t}\Bigr\rangle\, d{\bf x}.
\label{eq:5e}
\end{equation}
In the paper \cite{markov_NPA_2006} we have suggested the formula for emitted radiant power of soft fermionic excitations.
In particular, in the temporal gauge we have
\[
\langle Q_F\rangle= -\,\frac{1}{2}\,
\!\lim\limits_{\tau,\,V\rightarrow\infty} \frac{1}{\tau V}\!\int\limits_{-\tau/2}^{\tau/2}\!\!dt\!\int\limits_{V}\!d{\bf x}
\left\{\left\langle \frac{\partial\bar{\psi}_{\alpha}^i({\bf x},t)}{\partial t}\, \eta_{\alpha}^{\,i}({\bf x},t)\right\rangle +
\left\langle \bar{\eta}_{\alpha}^{\,i}({\bf x},t)\,
\frac{\partial\psi_{\alpha}^i({\bf x},t)}{\partial t}\right\rangle\right\}.
\]
This formula was defined in \cite{markov_NPA_2006} within semiclassical approximation and therefore the functions on the right-hand side
are classical Grassmann-valued ones. Comparing $\langle Q_F\rangle$ with (\ref{eq:5e}), we can draw the conclusion that if as stationary and homogeneous
fields $\xi^{(i)}$ we choose
\[
\xi^{(i)}({\bf x},t)=\left\{\bar{\psi}_{\alpha}^i({\bf x},t), \psi_{\alpha}^i({\bf x},t)\right\},
\]
then the volume field density of the generalized forces $f^{(i)}$ here will be
\[
f^{(i)}({\bf x},t)=\left\{\eta_{\alpha}^i({\bf x},t), -\bar{\eta}_{\alpha}^i({\bf x},t)\right\}.
\]
We believe that such a classification is valid both in (semi)classical and in quantum cases.\\
\indent
In the previous section for the generalized force, the fluctuation-dissipation relation
\begin{equation}
(\eta_{\alpha}^i\bar{\eta}_{\beta}^j)_{\omega{\bf q}}=
\frac{1}{2(2\pi)^4}\,i\hbar\,
\tanh\frac{1}{2}\beta(\hbar\omega-\mu)
\!\left\{G^{\,(R)\,ij}_{\alpha\beta}({\bf q},\omega)-G^{\,(A)\,ij}_{\alpha\beta}({\bf q},\omega)\right\}
\label{eq:5r}
\end{equation}
has been obtained. Here, the $(\eta_{\alpha}^i\bar{\eta}_{\beta}^j)_{\omega{\bf q}}$ function represents  Fourier component of the correlation function for fluctuations of source,
Eq.\,(\ref{eq:4p}). From the condition of mutuality between the generalized forces and coordinates in thermodynamically equilibrium system, a similar relation for soft part of the
total fermionic field of the system in question, is immediately follows:
\begin{equation}
(\psi_{\alpha}^i\bar{\psi}_{\beta}^j)_{\omega{\bf q}}=
\frac{1}{2(2\pi)^4}\,i\hbar\,
\tanh\frac{1}{2}\beta(\hbar\omega-\mu)
\!\left\{S^{\,(R)\,ij}_{\alpha\beta}({\bf q},\omega)-S^{\,(A)\,ij}_{\alpha\beta}({\bf q},\omega)\right\},
\label{eq:5t}
\end{equation}
where now the Green's functions $S^{\,(R,\,A)\,ij}_{\alpha\beta}$ are given by Eq.\,(\ref{eq:5q}). For global equilibrium
system the condition of mutuality suggests that these Green's functions are inverse to the functions $G^{\,(R,\,A)\,ij}_{\alpha\beta}$
on the right-hand side of (\ref{eq:5r}).\\
\indent
Further, we will be interested in the semiclassical (long-wavelength) limit of relation (\ref{eq:5t}). In this case the retarded and advanced propagators
should be chosen in the HTL-approxima\-tion according to the rule
\[
S^{\,(R,\,A)\,ij}(q)\rightarrow\frac{1}{\hbar}\,\delta^{ij}\,^{\ast}\!S^{\,(R,\,A)}(q).
\]
Here, the $1/\hbar$ factor has arisen owing to a choice of our initial definition of the Green's functions $S^{\,(R,\,A)}$.
Within the approximation in question the following useful relation
\[
\,^{\ast}\!S^{\,(A)}(q)=\gamma^0(\,^{\ast}\!S^{\,(R)}(q))^{\dagger}\gamma^0 = -\,^{\ast}\!S^{\,(R)}(-q)
\]
is valid.\\
\indent
Taking into account the expression above in the limit $\hbar\rightarrow 0$, the FDT (\ref{eq:5t}) results in
\begin{equation}
(\psi_{\alpha}^i\bar{\psi}_{\beta}^j)_{\omega {\bf q}}=
\frac{1}{2i(2\pi)^4}\,\delta^{ij}
\tanh\Bigl(\frac{1}{2}\,\beta\mu\Bigr)\!
\left\{\!\,^{\ast}\!S^{\,(R)}_{\alpha\beta}(q)\, + \,^{\ast}\!S^{\,(R)}_{\alpha\beta}(-q)\right\},
\label{eq:5y}
\end{equation}
where an explicit form of the quark propagator $\,^{\ast}\!S^{\,(R)}(q)$ is given by Eqs.\,(B.5)\,--\,(B.8).
Now it is necessary to consider the soft spinor fields $\psi_{\alpha}^i$ and $\bar{\psi}_{\beta}^j$ are classical
Grassmann-valued functions. From the relation (\ref{eq:5y}) it immediately follows that in order for
the spectral density for fluctuations of soft fermionic field to be non-zero (in the semiclassical approximation), the chemical potential $\mu$ of
the system under consideration {\it must} be non-zero\footnote{\,In quantum case (\ref{eq:5t}) a similar statement takes place for
the static limit $\omega\rightarrow 0$, if the exact propagators $S^{\,(R,\,A)}({\bf q},\omega)$ have no a singularity on
$\omega$ in this limit.}.\\
\indent
By using an explicit form of the retarded quark propagator, it is not difficult to obtain
\[
\,^{\ast}\!S^{\,(R)}(q)\, + \,^{\ast}\!S^{\,(R)}(-q)=
2i\Bigl\{
{\rm Im}(\,^{\ast}\!\triangle_{+}(q))h_{+}(\hat{\mathbf q})+
{\rm Im}(\,^{\ast}\!\triangle_{-}(q))h_{-}(\hat{\mathbf q})
\Bigr\}.
\]
Taking into account the relationship
\[
{\rm Im}(\,^{\ast}\!\triangle_{\pm}(q))=
{\rm Im}(\,^{\ast}\!\triangle_{\pm}^{-1}(q))^{\ast}\,
|\,^{\ast}\!\triangle_{\pm}(q)|^{\,2}
\]
and the fact that by virtue of (B.7) and (B.8) one has
\[
{\rm Im}(\,^{\ast}\!\triangle_{\pm}^{-1}(q))^{\ast}=
\frac{\pi\,\omega_0^2}{2\,|{\bf q}|}
\left(1\mp\frac{\omega}{|{\bf q}|}\right)\!\theta({\bf q}^2-\omega^2),
\]
we obtain the final form of the FDT relation
\begin{equation}
(\psi_{\alpha}^i\bar{\psi}_{\beta}^j)_{\omega {\bf q}}=
\frac{1}{2(2\pi)^3}\,\delta^{ij}\tanh\Bigl(\frac{1}{2}\,\beta\mu\Bigr)\frac{\omega_0^2}{2\,|{\bf q}|}
\label{eq:5u}
\end{equation}
\[
\times\!\left\{\left(1-\frac{\omega}{|{\bf q}|}\right)\!(h_{+}(\hat{\mathbf q}))_{\alpha\beta}|\,^{\ast}\!\triangle_{+}(q)|^{\,2}+
\left(1+\frac{\omega}{|{\bf q}|}\right)\!(h_{-}(\hat{\mathbf q}))_{\alpha\beta}|\,^{\ast}\!\triangle_{-}(q)|^{\,2}\right\}
\theta({\bf q}^2-\omega^2),
\]
instead of Eq.\,(\ref{eq:5y}).\\
\indent
Comparing the expression (\ref{eq:5u}) with that for the spectral density (\ref{eq:3t}) obtained within our simple model,
we see that they differ from each other by the $\tanh(\beta\mu/2)$ factor. Since the derivation of expression (\ref{eq:5u})
is more fundamental, it is natural, therefore, to assume that we have overlooked something in deriving (\ref{eq:3t}). Let us return
to the procedure of average in (\ref{eq:3e}). Careful analysis of (\ref{eq:3e}) points out that in this average the following
simple fact has not been taken into account, namely, the system under consideration is that with a varying number of particles and
antiparticles. In our case a baryon number $N$ that for massless thermal quarks and antiquarks can be presented as
\begin{equation}
N = N_f N_c\!\!\int\!\frac{d^{3}p}{(2\pi)^3}\,
\left\{
\frac{1}{{\rm e}^{\,\textstyle\beta(\varepsilon - \mu)} + 1}\; - \,
\frac{1}{{\rm e}^{\,\textstyle \beta(\varepsilon + \mu)} + 1}
\right\},
\label{eq:5i}
\end{equation}
(here, $\varepsilon = |{\bf p}|$), is a measure of change of the particle (and antiparticle) number. The spectral density (\ref{eq:3t}) is generally calculated at some fixed $N$.
To derive the correlation function $\left\langle\psi_{\alpha}^i(q)\bar{\psi}_{\beta}^j(-q^{\prime})\right\rangle$ considering a change of $N$, we use a trick
suggested by R.P. Feynman \cite{feynman_book_1972} in computing the free energy of the grand canonical ensemble.\\
\indent
Following \cite{feynman_book_1972}, we multiply the right-hand side of Eq.\,(\ref{eq:3e}) by some weighting factor of the type: $a_N {\rm e}^{-\beta \mu N}$. The weighting factor
accounts for different probabilities of different numbers of particles and antiparticles in the system. However, in our case as distinct from \cite{feynman_book_1972}, the baryon number
(\ref{eq:5i}) runs not only over all positive values, but also over all {\it negative} values by virtue of the above-mentioned possibility of exchange of the fermion number
between the hard and soft fermion subsystems of a quark-gluon plasma. This implies that instead of the factor ${\rm e}^{-\beta \mu N}$ we need to take
${\rm e}^{-\beta\mu\vert N \vert}$. Further, a careful distinction must be made between even and odd number of fermions. As is known
(see, for example, \cite{green_book_1987}) this is achieved by introducing the fermion number operator $(-1)^{\hat{F}}$ (the factor $(-1)^N$ in our case), which is identified
with the $a_N$ coefficient. Taking into account all the above-mentioned, instead of (\ref{eq:3e}) we have to write now
\[
\left\langle\psi_{\alpha}^i(q)\bar{\psi}_{\beta}^j(-q^{\prime})\right\rangle=
-\frac{\,g^2}{(2\pi)^6}\;\delta^{ij}\!\!
\sum\limits_{N=-\infty}^{+\,\infty}\!(-1)^{N}{\rm e}^{-\beta\mu|N|}\!\!\!
\sum\limits_{\;\zeta=\,\textmd{Q},\,\bar{\textmd{Q}}\,}\!\!\left(\frac{C_{\theta}^{(\zeta)}}{N_c}\right)
\!\int\!{\bf p}^2\!
\left[\,f_{|{\bf p}|}^{(\zeta)} + f_{|{\bf p}|}^{(\textmd{G})}\,\right]\!
\frac{d|\,{\bf p}|}{2\pi^2}
\]
\[
\times\!\int\!\frac{d\Omega_{{\bf v}}}{4\pi}
 \left(\!\,^{\ast}\!S(q)\chi\right)_{\alpha}
\left(\bar{\chi}\,^{\ast}\!S(-q^{\prime})\right)_{\beta}
\delta(\omega-{\bf v}\cdot{\bf q})\,\delta(q-q^{\prime}).
\]
Considering
\begin{equation}
\sum\limits_{N=-\infty}^{+\,\infty}\!(-1)^{N}{\rm e}^{-\beta\mu|N|}=
1 + 2\sum\limits_{N=1}^{+\,\infty}\!(-1)^{N}{\rm e}^{-\beta\mu N}
= \tanh\Bigl(\frac{1}{2}\,\beta\mu\Bigr),
\label{eq:5o}
\end{equation}
we conclude that to account for the fact of a change of the particle number in spectral density (\ref{eq:3t}), it is sufficient to multiply it by the factor $\tanh(\beta\mu/2)$.
This results in an identical coincidence between the spectral densities (\ref{eq:3t}) and (\ref{eq:5u}). The only difference is that for the first spectral density, the sum (\ref{eq:5o}) converges
only for $\mu > 0$, while (\ref{eq:5u}) is true for any values of the chemical potential.\\
\indent
In the case of $\mu < 0$ as the weighting factor it is necessary to take ${\rm e}^{+\beta\mu\vert N \vert}$, where now we have to set $a_N=-(-1)^N$. Then instead of the sum (\ref{eq:5o})
we will have
\[
-\!\!\sum\limits_{N=-\infty}^{+\,\infty}\!(-1)^{N}{\rm e}^{\,\beta\mu|N|}=
-1 - 2\sum\limits_{N=1}^{+\,\infty}\!(-1)^{N}{\rm e}^{\,\beta\mu N}
= \tanh\Bigl(\frac{1}{2}\,\beta\mu\Bigr)
\]
and in doing so we return again to the same factor, but with the negative chemical potential.

\section{\bf Conclusion}
\setcounter{equation}{0}

In this paper we have proved two independent derivations of the spectral density for soft fermionic excitations in an equilibrium quark-gluon plasma.
Careful consideration of all features that are common to excitations (both soft and hard) obeying Fermi-Dirac statistics, has allowed us to obtain total
coincidence of the final expressions. This in itself is a remarkable fact, since the initial premises of these two approaches, are supported on different physical
grounds. Another very intriguing feature is rather strong formal similarity of our first derivation with the bosonic case. The latter was considered in Section 2
for comparison.\\
\indent
The result obtained in this work appears to be sufficient justification for our pseudoclassical approach \cite{markov_NPA_2007, markov_IJMPA_2010}
in a description of the dynamics of hard and soft excitations of a hot QCD plasma obeying various statistics, at least on the level of thermal fluctuations
(at soft scale $gT$), within perturbation theory. The next step should be an extension of the model suggested in \cite{markov_NPA_2007, markov_IJMPA_2010}
to a description of highly excited states, i.e. QGP far from thermal equilibrium and beyond the limits of perturbation theory (strongly coupled plasma).

\section*{\bf Acknowledgments}
The authors are grateful to Prof. Alexander N. Vall and Andrey E. Radzhabov for useful discussions.
This work was supported by the Russian Foundation for Basic Research (project no 09-02-00749),
by the grant of the president of Russian Federation for the support of the leading
scientific schools (NSh-1027.2008.2), and in part by the Federal Target Programs “Development
of Scientific Potential in Higher Schools” (project 2.2.1.1/1483, 2.1.1/1539) and
“Research and Training Specialists in Innovative Russia, 2009–2013”, Contract 02.740.11.5154.

\newpage

\section*{\bf Appendix A}
\setcounter{equation}{0}

In Section 2 we make use the following formula of the integration with respect to commuting color charge $Q^a$:
$$
\int\!dQ\,Q^aQ^b\!=\left(\frac{C_2^{(\zeta)}}{d_A}\right)\!\delta^{ab},\quad \zeta=\textmd{Q},\bar{\textmd{Q}},\textmd{G},
\eqno{({\rm A}.1)}
$$
where as the integration measure $dQ$ the following expression is meant \cite{heinz_1983, heinz_1985, litim_2002}:
$$
dQ=d^{N_c}Q\,c_R\,\delta(Q^aQ^a-q_2)\ldots\,.
\eqno{({\rm A}.2)}
$$
The constant $c_R$ is fixed by the normalization condition $\int\!dQ=1$. With the constant $q_2$ one relates the so-called
quadratic Casimirs. The dots on the right-hand side of (A.2) denotes delta-functions ensuring the conservation of the
higher Casimirs for the color groups $SU(N_c),\,N_c\geq 3$.\\
\indent
By the quadratic Casimirs in quantum field theory, the numerical constants of the unit operator $\hat{I}$
of proper dimension on the right-hand side of the expressions
\[
\begin{split}
&t^at^a=C_F\hat{I},\\
&T^aT^a=C_A\hat{I},
\end{split}
\]
are commonly taken, where $(T^a)^{bc}=-if^{abc},\,C_F=(N_c^2-1)/2N_c$, and $C_A=N_c$. To provide the correct expressions in integrating
with respect to the color charges as the quark and gluon quadratic Casimirs in (A.1), we will mean
\[
C_2^{(\textmd{Q},\bar{\textmd{Q}})}\equiv{\rm tr}(t^at^a)=T_Fd_A,
\eqno{({\rm A}.3)}
\]
\[
C_2^{(\textmd{G})}\equiv{\rm tr}(T^aT^a) = C_Ad_A.
\eqno{({\rm A}.4)}
\]
Here, $d_A=N_c^2-1$ is dimension of the gauge group and $T_F\!=\!\frac{1}{2}$ is the index of the fundamental representation.
For such a choice of the constants $C_2^{(\zeta)}$, in particular, we derive the correct expression for the Debye screening mass
(\ref{eq:2g}). The constant $q_2$ in measure (A.2) is simply $C_2^{(\zeta)}$. This is exactly what is given in the paper \cite{litim_2002}.


\section*{\bf Appendix B}
\setcounter{equation}{0}

In this Appendix all necessary formulae for the gluon and quark propagators in the hard thermal loop
approximation and also some formulae for the integration over solid angle we use throughout this
work, are given.\\
\indent
The HTL-resummed gluon propagator in the covariant gauge is defined by the following expression:
$$
\,^{\ast}{\cal D}_{\mu \nu}(k)= -
P_{\mu \nu}(k) \,^{\ast}\!\Delta^t(k) -
Q_{\mu \nu}(k) \,^{\ast}\!\Delta^l(k) +
\xi D_{\mu \nu}(k)\Delta^0(k),
\eqno{({\rm B}.1)}
$$
where $\Delta^0(k)=1/k^2$; $\!\,^{\ast}\!\Delta^{t,\,l}(k) =
1/(k^2 - \Pi^{t,\,l}(k)), \,
\Pi^t(k) = \frac{1}{2} \Pi^{\mu \nu}(k) P_{\mu \nu}(k),$ and
$\Pi^l(k) = \Pi^{\mu\nu}(k)Q_{\mu\nu}(k)$
with the gluon self-energy
$\Pi_{\mu\nu}(k)$
$$
\Pi_{\mu\nu}(k)=m_D^2\left\{
u_{\mu}u_{\nu} - \omega\!
\int\frac{d\Omega_{\bf v}}{4\pi}\,
\frac{v_{\mu}v_{\nu}}{v\cdot k +i\epsilon}
\right\},
\eqno{({\rm B}.2)}
$$
$\xi$ is a gauge-fixing parameter. An explicit form of the Lorentz matrices
$P_{\mu\nu}(k)$, $Q_{\mu\nu}(k)$, and $D_{\mu \nu}(k)$ in the chosen gauge, is
$$
P_{\mu\nu}(k) = g_{\mu\nu}-D_{\mu\nu}(k)-Q_{\mu\nu}(k),\quad
Q_{\mu \nu}(k) =
\frac{\bar{u}_{\mu}(k)\bar{u}_{\nu}(k)}{\bar{u}^2(k)}\,,\quad
D_{\mu \nu}(k)=\frac{k_{\mu}k_{\nu}}{k^2}\,,\quad
\eqno{({\rm B}.3)}
$$
$$
\bar{u}_{\mu}(k)=k^2u_{\mu}-k_{\mu}(k\cdot u),
$$
where $u_{\mu}$ is the global four-velocity of a non-Abelian plasma. The Lorentz matrices
possess evident properties of ordinary projectors
$$
P^2=P,\quad Q^2=Q,\quad D^2=D,\quad
PQ=PD=QD=0.
\eqno{({\rm B}.4)}
$$
\indent
Further, the medium modified quark propagator $\,^{\ast}\!S(q)$ has the following form:
$$
\,^{\ast}\!S(q) = h_{+}(\hat{\mathbf q}) \,^{\ast}\!\triangle_{+}(q) +
h_{-}(\hat{\mathbf q}) \,^{\ast}\!\triangle_{-}(q),
\eqno{({\rm B}.5)}
$$
where the matrix functions
$$
h_{\pm}(\hat{\bf q})=(\gamma^0 \mp \hat{{\bf q}}\cdot\boldsymbol{\gamma}\,)/2
\eqno{({\rm B}.6)}
$$
with $\hat{{\bf q}} \equiv {\bf q}/\vert {\bf q} \vert$, are the spinor projectors
onto eigenstates of helicity, and
$$
\,^{\ast}\!\triangle_{\pm}(q) =
-\,\frac{1}{\omega\mp [\,\vert{\bf q}\vert + \delta\Sigma_{\pm}(q)]}
\eqno{({\rm B}.7)}
$$
are the `scalar' quark propagators, where in turn
$$
\delta \Sigma_{\pm}(q) =
\frac{\omega_0^2}{\vert {\mathbf q} \vert}
\biggl[\,1-\biggl(1\mp\frac{\vert {\mathbf q}
\vert }{\omega}\biggr) F\biggl(\frac{\omega}{\vert {\mathbf q}\vert}\biggr)\biggr]
\eqno{({\rm B}.8)}
$$
with
$$
F(z) = \frac{z}{2}\biggl[\,\ln\bigg{\vert}\frac{1 + z}{1 - z}\bigg{\vert}
-i\pi\theta(1-\vert z\vert)\biggr],
$$
are the scalar quark self-energies for normal $(+)$ and plasmino $(-)$ modes.
For the $h_{\pm}(\hat{\bf q})$ matrices the following identities
$$
h_{\pm}(\hat{\bf q})h_{\pm}(\hat{\bf q})=0,\quad
h_{\pm}(\hat{\bf q})h_{\mp}(\hat{\bf q})h_{\pm}(\hat{\bf q})=h_{\pm}(\hat{\bf q})
\eqno{({\rm B}.9)}
$$
are valid.\\
\indent
Finally, we give an explicit form of some simple integrals over solid angle $d\Omega_{\bf v}$:
\[
\begin{split}
&\int\frac{d\Omega_{\bf v}}{4\pi}\,\delta(\omega-{\bf v}\cdot{\bf k})=
\frac{1}{2|{\bf k}|}\,\theta({\bf k}^2-\omega^2),\\
&\int\frac{d\Omega_{\bf v}}{4\pi}\,v^i\delta(\omega-{\bf v}\cdot{\bf k})=
\frac{1}{2|{\bf k}|}\,\theta({\bf k}^2-\omega^2)\,\frac{\omega}{|{\bf k}|}\hat{k}^i,\\
&\int\frac{d\Omega_{\bf v}}{4\pi}\,v^iv^j\delta(\omega-{\bf v}\cdot{\bf k})=
\frac{1}{2|{\bf k}|}\,\theta({\bf k}^2-\omega^2)
\left\{\frac{{\bf k}^2-\omega^2}{2{\bf k}^2}(\delta^{ij}-\hat{k}^i\hat{k}^j)+
\frac{\omega^2}{{\bf k}^2}\hat{k}^i\hat{k}^j\right\},
\end{split}
\tag{{\rm B}.10}
\]
where $\hat{k}^i\equiv k^i/|{\bf k}|$.


\section*{\bf Appendix C}
\setcounter{equation}{0}

In this Appendix we discuss a problem of construction of the integration measure with respect to anticommuting color charges $\theta^i$ and
$\theta^{\dagger i}$. These charges obey the following dynamical equations \cite{barducii_1977}:
\[
\begin{split}
&\frac{d\theta^i(t)}{dt} + igv^{\mu}A^a_{\mu}(t,{\bf v}t)(t^a)^{ij}
\theta^j(t)=0, \quad\;
\left.\theta^i_0=\theta^i(t)\right|_{\,t=0}\\
&\frac{d\theta^{\dagger\, i}(t)}{dt} - igv^{\mu}A^a_{\mu}(t,{\bf v}t)
\theta^{\dagger j}(t)(t^a)^{ji}=0, \quad
\left.\theta^{\dagger\, i}_0=\theta^{\dagger\, i}(t)\right|_{\,t=0},
\end{split}
\tag{{\rm C}.1}
\]
where $v^{\mu}=(1,{\bf v})$. Here, we have neglected the interaction of the Grassmann charges with soft $\psi$-fields induced  by thermal fluctuations
in a hot non-Abelian plasma. This system of equations permits an quadratic integral of motion
$$
\theta^{\dagger i}(t)\theta^i(t)=\theta_0^{\dagger i}\theta_0^i
\equiv C_{\theta}^{(\zeta)},\quad \zeta=\textmd{Q},\bar{\textmd{Q}},\textmd{G}.
$$
The integral of motion is analog of the invariant $Q^a(t)Q^a(t)=Q^a_0Q^a_0\,(\,\equiv C_2^{(\zeta)})$ given in Appendix A
and it as well as the invariant $C_2^{(\zeta)}$ depends on the type of hard parton to which the Grassmann color charges concern.\\
\indent
As a definition of the Grassmann color charge integration measure we put the following expression:
$$
d\theta d\theta^{\dagger}\equiv
\left(\prod\limits_{i=1}^{N_c}d\theta^i d\theta^{\dagger i}\right)
\!f(\theta^{\dagger}\theta),
\eqno{({\rm C}.2)}
$$
where $f(\theta^{\dagger}\theta)$ is unknown for the time being function of the invariant
$\theta^{\dagger}\theta\equiv\theta^{\dagger i}\theta^i$. This measure is invariant with respect to
the involution $\dagger$. The function $f$ must formally play a role of a Dirac $\delta$-function
as it takes place, for example, in the definition of measure for the usual color charges, Eq.(A.2), i.e.,
$$
f(\theta^{\dagger}\theta)\sim\delta(\theta^{\dagger}\theta - C_{\theta}^{(\zeta)}).
$$
Let us look for an explicit form of this function in a way of a (finite) expansion in powers of $\theta^{\dagger}\theta$
$$
f(\theta^{\dagger}\theta)=f_0 + f_1\,\theta^{\dagger}\theta + f_2(\theta^{\dagger}\theta)^2
+\ldots+ f_{N_c-1}(\theta^{\dagger}\theta)^{N_c-1} + f_{N_c}(\theta^{\dagger}\theta)^{N_c}.
\eqno{({\rm C}.3)}
$$
Here, $f_n,\;n=0,1,\ldots,N_c$ are some constants depending on the group invariants. Below we
use the standard rules for the integration over the Grassmann variables
\[
\begin{split}
&\int\!d\theta^i=\!\int\!d\theta^{\dagger i}=0,\\
&\int\!d\theta^i\theta^i=\!\int\!\theta^{\dagger i}d\theta^{\dagger i}=1\;(\mbox{\it not summation!}).
\end{split}
\]
\indent
Let us require that for each function of the form $g = g(\theta^{\dagger}\theta)$ the relation
$$
\int\!d\theta d\theta^{\dagger} g(\theta^{\dagger}\theta) = g(C_{\theta}^{(\zeta)}).
\eqno{({\rm C}.4)}
$$
was fulfilled. The function $g(\theta^{\dagger}\theta)$ represents a finite polynomial in $\theta^{\dagger} \theta$ similar to (C.3). Making use the integration rules over
the Grassmann variables, it is not difficult to obtain that the relation (C.4) will be fulfilled if the function $f(\theta^{\dagger}\theta)$ has the form
$$
f(\theta^{\dagger}\theta)=\frac{1}{N_c!}\,\Bigl\{\bigl(C_{\theta}^{(\zeta)}\bigr)^{N_c}\! + \bigl(C_{\theta}^{(\zeta)}\bigr)^{N_c-1}\theta^{\dagger}\theta + \,\ldots\, +
C_{\theta}^{(\zeta)}(\theta^{\dagger}\theta)^{N_c-1} + (\theta^{\dagger}\theta)^{N_c\!}\Bigr\}.
\eqno{({\rm C}.5)}
$$
As the simplest consequences of this fact we have the normalization condition
$$
\int\!d\theta d\theta^{\dagger}=1.
$$
and also the integral
$$
\int\!d\theta d\theta^{\dagger}\,\theta^{\dagger i}\theta^j=
\left(\frac{C_{\theta}^{(\zeta)}}{N_c}\right)\!\delta^{ij}.
\eqno{({\rm C}.6)}
$$
It is worth noting here that the integral (C.6) enables us to calculate the other more complicated expressions involving the anticommuting color charges of different particles.
So when doing the calculation of the probability of soft quark bremsstrahlung \cite{markov_IJMPA_2010}, we are faced with a color structure of the following type:
$$
C_{\theta\theta}^{(1;2)}\equiv
(\theta_{1}^{\dagger}t^a\theta_{2})(\theta_{2}^{\dagger}t^a\theta_{1}).
$$
Here, the labels 1 and 2 concern to the first and second hard particles involved in the scattering process in question.
The integration over the Grassmann color charges of these two hard partons of this color factor is easily carried out by
the general formula (C.6)
$$
\int\!d\theta_1 d\theta^{\dagger}_1\! \int\!\!d\theta_2 d\theta^{\dagger}_2
\,C_{\theta\theta}^{(1;2)}=
-{\rm tr}\,(t^at^a)\left(\frac{C_{\theta}^{(1)}}{N_c}\right)\!
\left(\frac{C_{\theta}^{(2)}}{N_c}\right)
=-C_F\!\left(\frac{C_{\theta}^{(1)}C_{\theta}^{(2)}}{N_c}\right).
$$
The explicit value obtained for the color factor $C_{\theta \theta}^{(1;2)}$, in exact coincides with a similar one
in the paper \cite{markov_IJMPA_2010} derived from a basically different reasoning by means of much more cumbersome calculations.\\
\indent
It is also easy to calculate more nontrivial expression of the type $\int\!d\theta d\theta^{\dagger}\,\theta^{\dagger i}\theta^j\theta^{\dagger k}\theta^l$.
Obviously, the given expression has to be antisymmetric with respect to a permutation of indices
$i\rightleftharpoons k$ and $j\rightleftharpoons l$. Therefore, we have
$$
\int\!\!d\theta d\theta^{\dagger}\,\theta^{\dagger i}\theta^j\theta^{\dagger k}\theta^l =
a(\delta^{ij}\delta^{kl}-\delta^{il}\delta^{kj}),
\eqno{({\rm C}.7)}
$$
where $a$ is some constant. A rather easy calculation leads to
$$
a=\frac{\bigl(C_{\theta}^{(\zeta)}\bigr)^{2}}{N_c(N_c-1)}.
$$
The integral
$$
\int\!\!d\theta d\theta^{\dagger}(\theta^{\dagger}t^a\theta)(\theta^{\dagger}t^b\theta)=
\frac{\bigl(C_{\theta}^{(\zeta)}\bigr)^{2}}{N_c(N_c-1)}\,
\bigl[\,{\rm tr}\,t^a{\rm tr}\,t^b - {\rm tr}\,t^at^b\bigr]=
-\frac{\;T_F\bigl(C_{\theta}^{(\zeta)}\bigr)^{2}}{N_c(N_c-1)}\;\delta^{ab}
$$
is a particular consequence of relation (C.7). In a similar way we can calculate the other more complicated integrals.\\
\indent
Thus, setting into consideration the integration over the Grassmann color charges enables us to automatize fully
the procedure of calculation of various color factors which appear when doing the determination of the scattering
probabilities involving hard and soft fermionic excitations. In doing so, we practically achieve here, perfect analogy
with the purely bosonic case \cite{markov_AOP_2004_2005}.

\end{document}